\documentclass[aps,pra,showpacs,twocolumn,superscriptaddress]{revtex4}

\usepackage{graphicx}
\usepackage{dcolumn}
\usepackage{bm}
\usepackage {amsmath}

\begin{document}

\title{Nonclasscial interference between independent intrinsically pure single photons at telecom wavelength}


\author{Rui-Bo Jin}
\email{ruibo@nict.go.jp}
\affiliation{Advanced ICT Research Institute, National Institute of Information and Communications Technology (NICT), 4-2-1 Nukui-Kitamachi, Koganei, Tokyo 184-8795, Japan}
\author{Kentaro Wakui}
\affiliation{Advanced ICT Research Institute, National Institute of Information and Communications Technology (NICT), 4-2-1 Nukui-Kitamachi, Koganei, Tokyo 184-8795, Japan}
\author{Ryosuke Shimizu}
\email{r-simizu@pc.uec.ac.jp}
\affiliation{Center for Frontier Science and Engineering, University of Electro-Communications, Tokyo 182-8585, Japan}
\author{Hugo Benichi}
\affiliation{Advanced ICT Research Institute, National Institute of Information and Communications Technology (NICT), 4-2-1 Nukui-Kitamachi, Koganei, Tokyo 184-8795, Japan}
\author{Shigehito Miki}
\affiliation{Advanced ICT Research Institute, National Institute of Information and Communications Technology (NICT), \\ 588-2 Iwaoka, Kobe 651-2492, Japan}
\author{Taro Yamashita}
\affiliation{Advanced ICT Research Institute, National Institute of Information and Communications Technology (NICT), \\ 588-2 Iwaoka, Kobe 651-2492, Japan}
\author{Hirotaka Terai}
\affiliation{Advanced ICT Research Institute, National Institute of Information and Communications Technology (NICT), \\ 588-2 Iwaoka, Kobe 651-2492, Japan}
\author{Zhen Wang}
\affiliation{Advanced ICT Research Institute, National Institute of Information and Communications Technology (NICT), \\ 588-2 Iwaoka, Kobe 651-2492, Japan}
\author{Mikio Fujiwara}
\affiliation{Advanced ICT Research Institute, National Institute of Information and Communications Technology (NICT), 4-2-1 Nukui-Kitamachi, Koganei, Tokyo 184-8795, Japan}
\author{Masahide Sasaki}
\affiliation{Advanced ICT Research Institute, National Institute of Information and Communications Technology (NICT), 4-2-1 Nukui-Kitamachi, Koganei, Tokyo 184-8795, Japan}

\date{\today }

\begin{abstract}
We demonstrate a Hong-Ou-Mandel interference between two independent, intrinsically pure,  heralded single photons from spontaneous parametric down conversion (SPDC) at telecom wavelength.
A visibility of $85.5\pm8.3\%$ was achieved without using any bandpass filter.
Thanks to the group-velocity-matched SPDC and superconducting nanowire single photon detectors (SNSPDs), the 4-fold coincidence counts are  one order higher than that in the previous experiments.
%
The combination of bright single photon sources and SNSPDs is a crucial step for  future practical quantum info-communication systems at telecom wavelength.

\end{abstract}



\maketitle

\section{Introduction}
Nonclassical interferences between independent single photons
(NIBISP) play an important role in quantum information processing.
Four folds NIBISP have been realized at near-infrared (NIR) wavelength using crystals \cite{Kaltenbaek2006, Mosley2008a, Tanida2012} and optical fibers \cite{Soller2011},  and at telecom wavelength using crystals \cite{Xue2010}, fibers \cite{Takesue2007}  and  silicon wire waveguides \cite{Harada2011, Takesue2012}.
However, as shown in Ref \cite{Xue2010, Takesue2007, Harada2011}, NIBISP at telecom wavelength are suffering from two problems: the less efficient photon sources and the low-performance detectors.
In the traditional photon sources based on spontaneous parametric down conversion (SPDC), the signal and idler
photons have correlated frequencies, i.e., the sources are not spectrally pure.
To increase the purity, we need to employ narrow bandpass filters.
This method, however, severely decreases the count rate of the photon sources.
Concerning the detectors at telecom wavelength, nowadays the commercially available and widely used single photon detectors  are avalanched photon detectors (APD) \cite{id210}.
This kind of detectors suffer from long dead times and high after-pulses.
Typically, the APDs are operated with a dead time of about 10 $\mu$s for reasonable low dark counts and after pulses \cite{id210}.
10 $\mu$s corresponds to a maximum count rate of less than 100 kilo counts per second (Kcps).
As a result, in previous experiments \cite{Evans2010, Yabuno2012} researchers employed pulse pickers to decrease the laser repetition rate, e.g., from 76 MHz down to 4.75 MHz \cite{Evans2010}, which decreased the count rates by 16 times.

To solve the first problem,  we utilize a recently developed intrinsically pure single photon source based on  SPDC in a periodically poled $\mathrm{KTiOPO_4}$
(PPKTP) crystal  with the group-velocity matching  condition.
PPKTP crystal with group-velocity matching  condition was first experimentally demonstrated by K{\"o}nig and Wong for second-harmonic generation \cite{Konig2004}.
Later, this condition was applied to SPDC for intrinsically pure photon state generation, in which the signal and idler photons from SPDC have no spectral correlation.
Therefore, there is no need to employ  bandpass filters to obtain  heralded single photons with high purity.
Consequently, such photon sources are more efficient than the traditional sources and have showed high brightness in experiments \cite{Evans2010, Gerrits2011, Eckstein2011, Grice2012, Yabuno2012}.
Recently, we showed that such sources had a very wide spectral tunability \cite{Jin2013a}.
The  spectral purity can be kept higher than 0.81 as the wavelength is tuned from 1460 nm to 1675 nm, which covers the S-, C-, L-, and U-band at telecommunication wavelengths.

To overcome the second problem, we employ  superconducting nanowire single photon detectors (SNSPDs) \cite{Miki2010},  operating in a free-running mode and  having a maximum detection efficiency of 23\%  with dark counts less than 100 cps at 1550 nm.
The dead time of the SNSPDs is about 30 ns, and the after pulses are almost negligible.
The  count rate of the SNSPDs can be as high as 30 Mcps.
This kind of high-performance detectors have been used in many  applications in quantum information processing \cite{Gerrits2011, Ikuta2013}.

In this paper, by  combining the bright single photon sources and SNSPDs, we demonstrate a Hong-Ou-Mandel interference between two independent and spectrally pure photon sources at telecom wavelength.
A visibility of $85.5\pm8.3\%$ was achieved at the pump power of 10 mW  with no use of any bandpass filter.
Thanks to the high performance of SNSPDs, we can obtain a coincidence count of 31 Kcps at the pump power of 100 mW.
This experiment opens the way for the future demonstration and practical  implementation of quantum information and communication protocols at telecom wavelength.

\section{Experiment  }

\begin{figure}[tbp]
\includegraphics[width= 0.45 \textwidth]{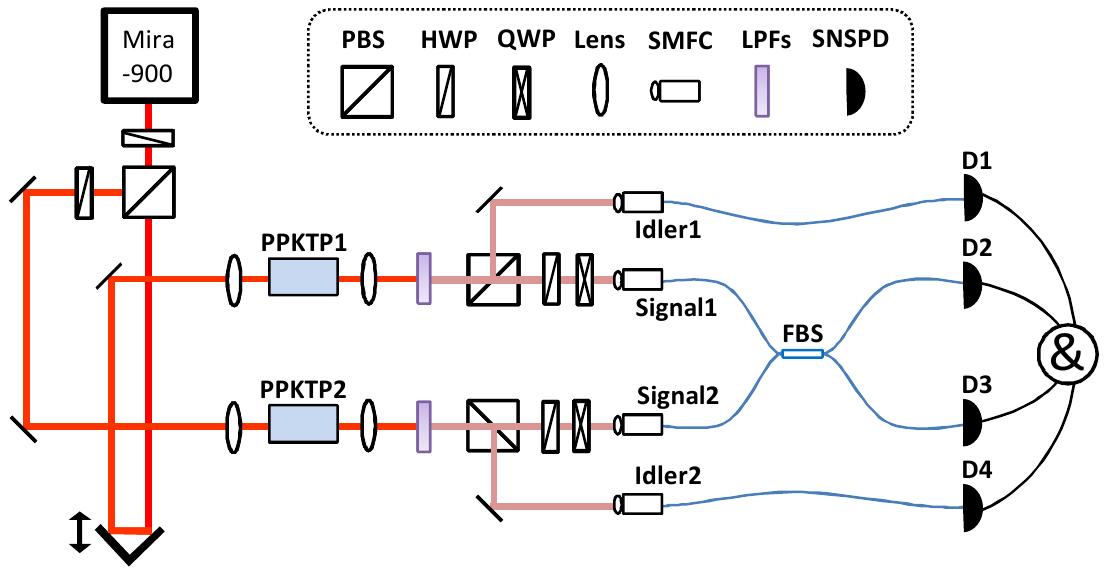}
\caption{ (Color online) The experimental setup. PBS (polarizing beam splitter), HWP (half wave plate), QWP (quarter wave plate),
          SMFC (single mode fiber coupler), LPFs (long-wave pass filters),  SNSPD (superconducting nanowire single photon detector), FBS (fiber beam splitter),  \& (coincidence counter).} \label{setup}
\end{figure}

The experimental setup is shown in Fig.\,\ref{setup}.
Picosecond laser pulses (76 MHz, 792 nm, temporal duration $\sim$ 2 ps) from a mode-locked Titanium sapphire laser (Mira900, Coherent Inc.) were divided into two paths and focused on two  30-mm-long PPKTP crystals with a poling period of 46.1 $\mu$m for type-II group-velocity-matched SPDC.
The temperatures of PPKTP1 and PPKTP2 were maintained at $28.8\,^{\circ}\mathrm{C}$ and  $32.5\,^{\circ}\mathrm{C}$ respectively, so as to achieve degenerate wavelength at 1584 nm.
The down-converted photons, i.e., the signal (H polarization) and idler (V polarization) were filtered by longpass filters and then separated by  polarizing beam splitters (PBS).
The signals were collected into a fiber beam splitter (FBS), while the idlers were collected into two single mode fibers (SMFs).
Finally, all the collected photons were sent to four SNSPDs, which were connected to a coincidence counter.

These SNSPDs have a design improved  from Ref  \cite{Miki2010}, using 4-mm-thick and 80-nm-wide niobium nitride (NbN) meander nanowire on 0.4-mm-thick  Si or MgO substrates.
The nanowire covers an area of 15 $\mu$m $ \times$ 15 $\mu$m.
The SNSPDs are installed in a Gifford-McMahon cryocooler system and are cooled to 2.1 K.
The detection efficiency is 23\%, 22\%, 19\%, 11\% at 1550 nm on detector D1, D2, D3 and D4, respectively.
The detection efficiency of these SNSPDs has a linearity as high as 30 MHz, which implies that  the SNSPDs can have a  single count rate  as high as 30 Mcps, much higher than the commercial APD systems in Ref \cite{id210}.
The dark counts can be  less than 100 cps by adjusting the bias current of the nanowires.
The detection efficiency of SNSPDs is polarization dependent. To improve the detection efficiency, we twist the fibers to compensate the polarization.

\begin{figure}[tbp]
\includegraphics[width= 0.4 \textwidth]{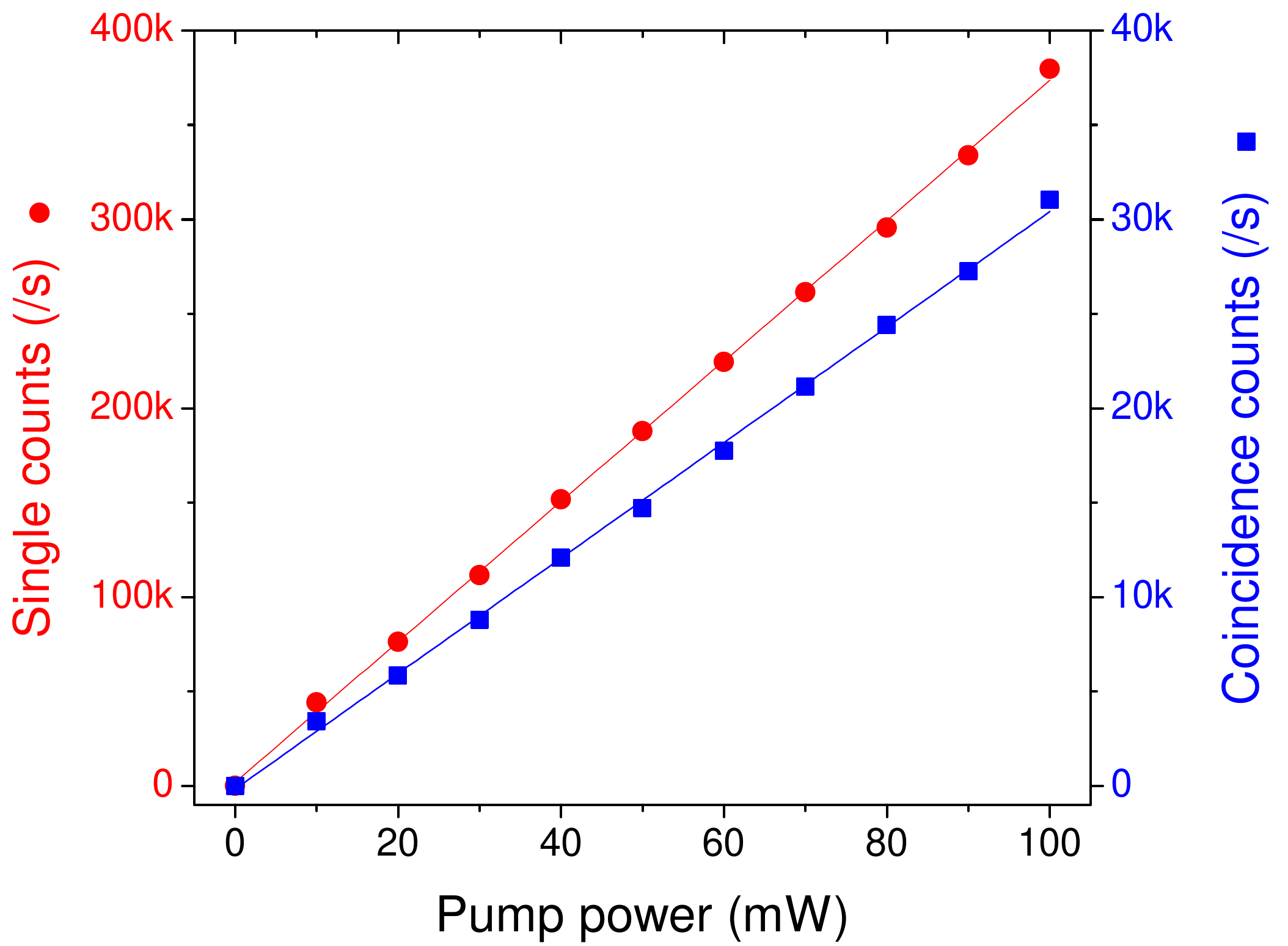}
\caption{ (Color online) Single counts (of idler1) and coincidence  counts (of signal1 and idler1) as a function of pump power.} \label{sccp}
\end{figure}

Firstly, we measured the single counts (of idler1) and coincidence counts (of signal1 and idler1) as a function of the pump power, as shown in Fig.\,\ref{sccp}.
Both the single counts and coincidence counts showed good linearity when the power was increased up to 100 mW in Fig.\,\ref{sccp}.
At 100 mW pump, the single count and coincidence count were 380 Kcps and 31 Kcps, which correspond to 0.06 pairs per pulse.
As far as we know, this coincidence count was the highest ever reported at telecom wavelength.
This coincidence can still be improved by increasing the pump power.

\begin{figure}[tbp]
\includegraphics[width= 0.3 \textwidth]{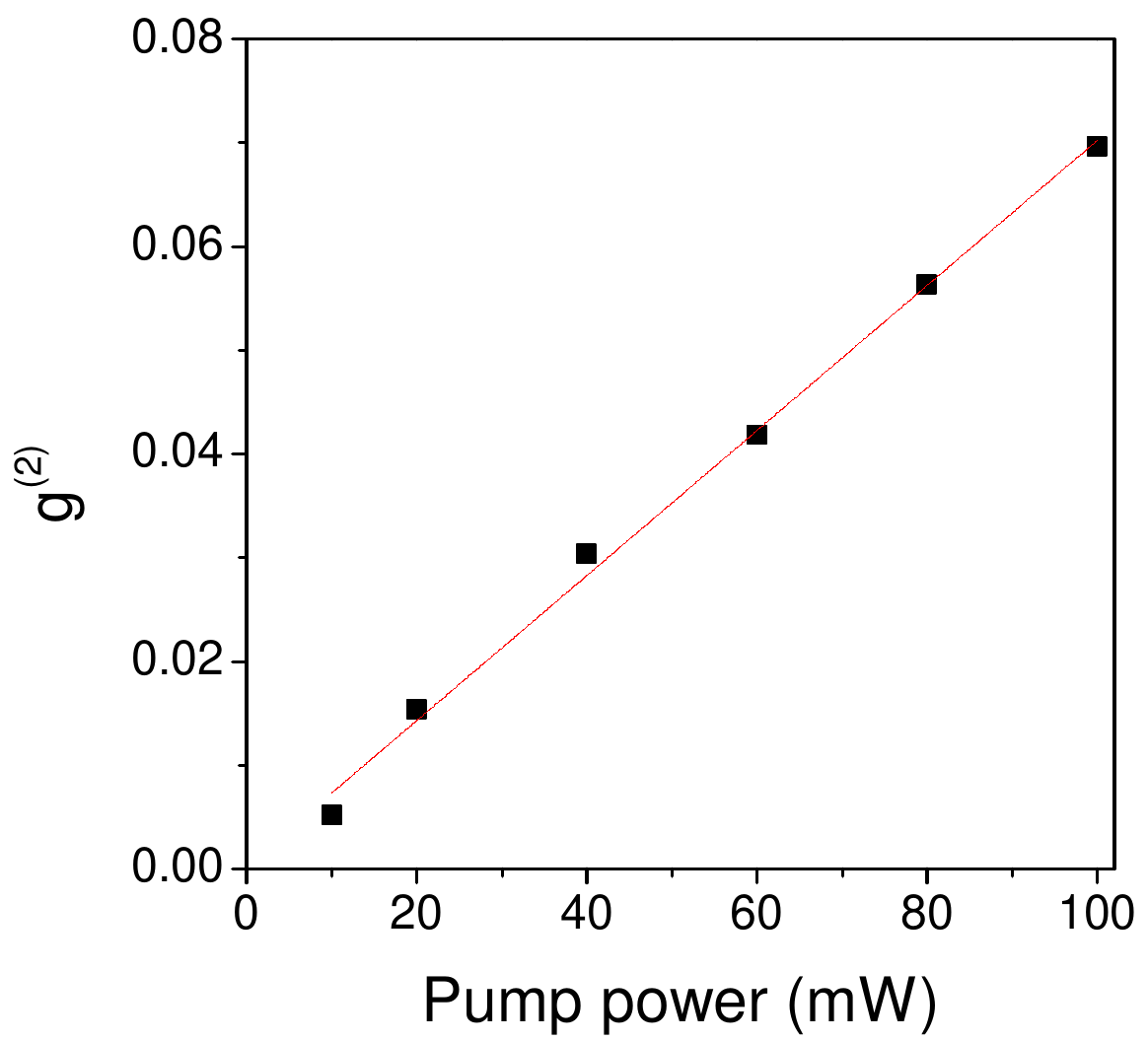}
\caption{ (Color online) $g^{(2)}$  as a function of pump power.} \label{g2}
\end{figure}

Then, we measured the second order coherence function $g^{(2)}$  with PPKTP1  using the equation \cite{Mosley2008b}
$$
 g^{(2)}  = \frac{{\left\langle {\hat a^\dag  \hat a^\dag  \hat a\hat a} \right\rangle }}{{\left\langle {\hat a^\dag  \hat a} \right\rangle ^2 }} \approx \frac{{2CC_{123}  \times SC_1 }}{{(CC_{12}  + CC_{13} )^2}},
$$
where, $SC_1$ is the single counts of D1; $CC_{12}$ and $CC_{13}$ are the coincidence counts of D1 and D2, D1 and D3, respectively;  $CC_{123}$ is the three folds coincidence counts of D1, D2 and D3.
Fig.\,\ref{g2} shows the value of  $g^{(2)}$ as a function of the pump power.
At  pump power of 100 mW, the  $g^{(2)}$ value is 0.07.   This low $g^{(2)}$  shows our source is close to a single photon source.

\begin{figure}[tbp]
\includegraphics[width= 0.35 \textwidth]{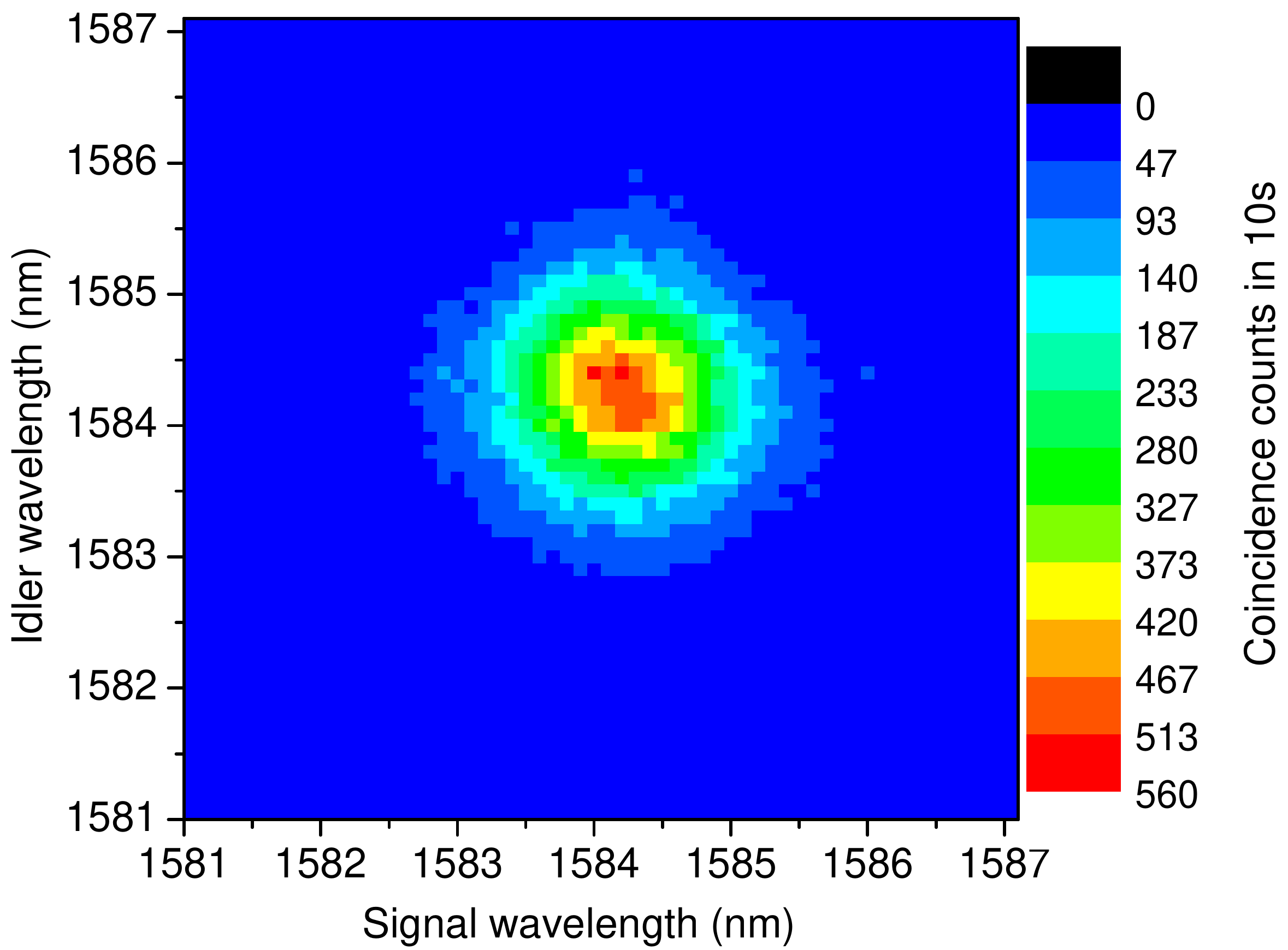}
\caption{ (Color online) Joint spectral intensity of the photon source. The coincidence counts between the signal and idler were accumulated for 10s for each point.} \label{jsi}
\end{figure}

To achieve a high  interference visibility, we need to make sure the heralded single photons from PPKTP1 and PPKTP2 are highly indistinguishable in every degree of freedom, e.g., spectral, temporal, spatial and in polarization.
To check the spectral purity of the photon source, we also measured the joint spectral intensity (JSI) of this source, as shown in Fig.\,\ref{jsi} \cite{Jin2013a}.
The calculated Schmidt number \cite{Eberly2006} of the JSI is 1.02, corresponding to a spectral purity of 0.82 \cite{Jin2013a}, which shows the high spectral indistinguishability of the signal and idler photons in this source.
The polarization indistinguishability of signal1 and signal2 is achieved by adjusting two sets of HWPs and QWPs after the PBSs.
The spacial indistinguishability is guaranteed by coupling all the photons into single mode fibers.

\begin{figure}[tbp]
\includegraphics[width= 0.45 \textwidth]{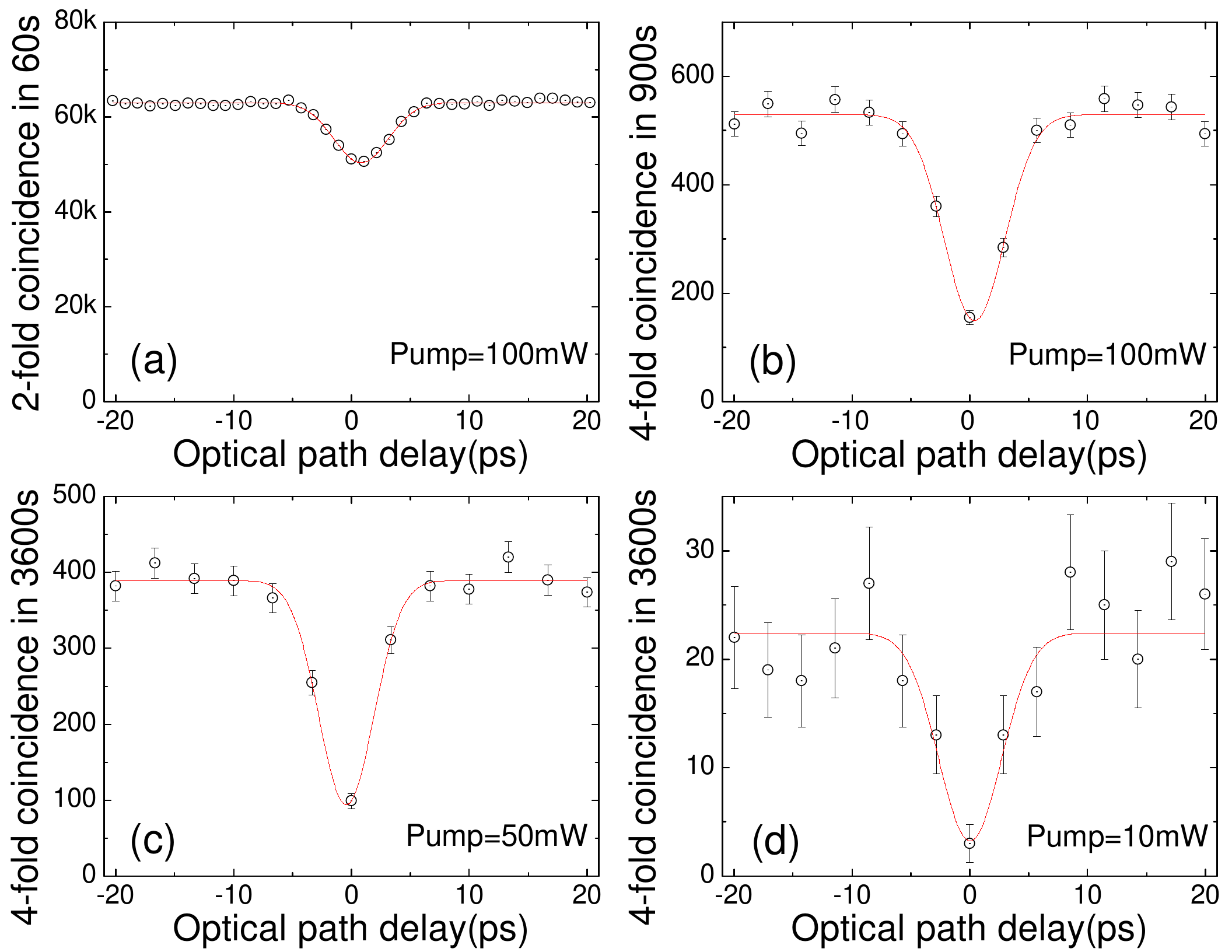}
\caption{ (Color online) Experimental results of two-photon interference from independent sources.
         The solid lines represent Gaussian fits to the data points.
         Error bars are equal to the square root of each data point, assuming Poissonian counting statistics.
        (a) shows a 2-fold coincidence counts of  D2 and D3 with a pump power of 100 mW for each PPKTP crystal, accumulated over 60s.
        The visibility is  $20.1\pm 0.3\%$ without background correction.
        (b), (c) and  (d) are  4-fold coincidence counts of  D1, D2, D3 and D4  with a pump power  of 100 mW, 50 mW and 10 mW  for each PPKTP crystal, respectively.
        The accumulation times are 900s, 3600s and 3600s.
        After correcting the background counts, the visibilities of (b), (c) and (d) were $71.8\pm2.7\%$, $75.8\pm2.4\%$ and $85.5\pm8.3\%$, respectively.
         } \label{result1}
\end{figure}

In order to check the temporal overlap of photons from signal1 and signal2, we  carried out a 2-fold Hong-Ou-Mandel interference \cite{Hong1987} between   them.
We scanned the optical path delay and recorded the coincidence counts, with a pump power of 100 mW for each PPKTP crystal.
The result is shown in Fig.\,\ref{result1} (a), with  a FWHM ( full width at half maximum ) of $5.4\pm0.14$ ps and visibility of $20.1\pm 0.3\%$, without background correction.
This is an interference between two thermal states, since without triggering by the idlers, the emission statistics of the signals is equivalent to light emitted by a thermal source \cite{Ou1999}.

After we knew the center point of the Hong-Ou-Mandel dip from  Fig.\,\ref{result1} (a),  we performed a 4-fold  Hong-Ou-Mandel interference between signal1 and signal2, with idler1 and idler2 as the triggers, with a pump power of 100 mW for each PPKTP crystal.
Fig.\,\ref{result1} (b) shows the results of such an interference, with FWHM of $6.0\pm0.4$ ps.
Without correcting the background, the raw visibility amounts to $58.2\pm 3.5\%$, which is higher than the 50\% classical limit \cite{Ou2007}.

The SPDC photon source not only emits one but also two or more pairs of photons, especially in the case of a high pump power.
The multi-photon components will lead to a 4-fold coincidence count, e.g., when signal1 and idler1 each has two photons as a result of double-pair emission from PPKTP1, while signal2 and idler2 each has one photon due to single-pair emission from PPKTP2.
In this case, the background counts are not only due to imperfect single photon interference but also caused by multiple photons coming from one of the sources.
To measure the background caused by these  multiple photons, we blocked only signal1 and measured a 4-fold coincidence counts of D1-D4, then only blocked signal2, and measured a 4-fold coincidence count of D1-D4.
The sum of these two sets of coincidence counts was the background count.
We obtained on average of 125 background counts for 900s in this measurement.
After correcting the background count, we obtained a visibility of $71.8\pm2.7\%$.

Since the background counts are mainly due to the multi-photon components in each source, we can decrease the pump power to reduce multi-photon emission, so as to improve the visibility.
Therefore, we reduced the pump power from 100 mW to 50 mW and 10 mW for each PPKTP and measured the coincidence counts again.
We obtained visibilities of $75.8\pm2.4\%$ and $85.5\pm8.3\%$ in the cases of  50 mW and 10 mW pump powers, respectively, as shown in Fig.\,\ref{result1} (c) and (d).
The background counts were subtracted using the same method.
From Fig.\,\ref{result1} (b-d), we noticed that the interference visibilities increased when the pump was decreased.

In the ideal condition, such Hong-Ou-Mandel interference of two heralded single photons should achieve a visibility of 100\%.
The degradation in our scheme are mainly caused by the imperfect spectral purity of the photon sources.
The relationship between visibility, purity and indistinguishability in such interference experiment can be proved as \cite{Mosley2008a, Mosley2008b, Osorio2013}
\begin{equation}\label{eq1}
V = \text{Tr}[\rho _1 \rho _2 ] = \frac{{ \text{Tr}[\rho _1^2 ] +  \text{Tr}[\rho _2^2 ] - \left\| {\rho _1  - \rho _2 } \right\|^2 }}{2},
\end{equation}
where V is the interference visibility; $\rho _1$  and $\rho _2$ are the density matrices of signal1 and signal2 (in Fig.\,\ref{setup});
$ \text{Tr}[\rho _1^2 ]$ and  $ \text{Tr}[\rho _2^2 ]$ are the purities of  the two signal photons;
$\left\| {\rho _1  - \rho _2 } \right\|^2 $ is the indistinguishability between them, and $\left\| \rho  \right\|^2 {\rm{ = }} \text{Tr}[\rho ^\dag  \rho ] $.
In our experiment, the measured spectral purities of signal1 and signal2 were 0.82, as shown in Fig.\,\ref{jsi}.
The spectral indistinguishability between signal1 and signal2 can be roughly estimated as 0, since these two PPKTP crystals have identical design and the measured spectra of  the two signal photons are almost the same, so the density matrices of signal1 and signal2 in the degree of spectrum freedom should be almost the same.
Some residual distinguishabilities in other  degree of  freedom between the two photons may also decrease the interference visibility, e.g., distinguishabilities caused by the slightly asymmetric ratio of the FBS, the imperfect polarization compensation for the FBS.
Therefore,  from Eq.\,\ref{eq1} we can estimate the upper bound of the visibility as $82\%$, which corresponds well with our measured visibility of $85.5\pm8.3\%$.
This visibility is still higher than that in the previous experiments at telecom wavelength \cite{Xue2010, Takesue2007, Harada2011}.

\section{Discussion}

Many fundamental multi-photon counting experiments have been done in the NIR wavelength range (around 800 nm), where bright sources and efficient, low-noise detectors are available.
In contrast, similar experiments in the telecom wavelength region are still difficult because of limited performance of sources and detectors.
Our source and detectors may change such a situation.

It is noteworthy to compare the brightness of our experiment with the previous schemes for quantum interference between independent single sources from SPDCs.
In Refs. \cite{Kaltenbaek2006, Mosley2008a, Tanida2012} in the NIR range,  the 4-fold coincidence counts were about 0.17 cps, 0.27 cps and 0.45 cps.
In Refs. \cite{Xue2010, Takesue2007, Harada2011} in the telecom range, the 4-fold coincidence counts were about 0.015 cps, 0.060 and 0.011cps.
In our scheme, the 4 fold coincidence counts is 0.58 cps at the pump power of 100 mW.
So, our count rate in the telecom range is in the same order and even a little higher than the traditional photon count rates in the NIR range,
and one order higher than  previous experiments at the telecom wavelengths.

Our system has the potential to achieve a brightness of the same level as in Refs \cite{Yin2012, Ma2012} by  increasing the pump power.
As a next step our sources and the SNSPDs will be applied for practical field tests of quantum teleportation, entanglement swapping, and so on, not only with optical fiber networks but also in free space.
Actually, the telecom wavelengths are eye safe and have less air absorption than the NIR wavelength regions.

\section{Conclusion}
In summary, we have demonstrated a quantum interference between two independent sources at telecom wavelength with two spectrally pure photon sources and SNSPDs.
We have achieved  coincidence counts of 31 Kcps at 100 mW pump, and  a visibility of $85.5\pm8.3\%$ at 10 mW pump.
This experiment shows that the photon counting experiment at the telecom wavelength range can achieve the same count rates as that in the  NIR range.
Our scheme is useful for practical quantum information and communication systems.

\section*{Acknowledgements}
The authors are grateful to  T. Gerrits for helpful discussion.
This work was supported by the Founding Program for World-Leading Innovative R\&D on Science and Technology (FIRST).

\end{document}